\documentclass[nonacm,sigplan]{acmart}

\usepackage[normalem]{ulem}
\usepackage[frozencache,cachedir=.]{minted}
\usepackage{tikz}
\usepackage{enumitem}
\usepackage{array}
\usepackage{booktabs}
\usepackage{cleveref}
\usepackage{comment}
\usepackage{xspace}
\usepackage{listings}
\usepackage{svg}
\usepackage{epigraph}
\usepackage{fancyvrb}
\usepackage{caption}
\usepackage{subcaption}
\usepackage{tikz}
\usepackage{tikzpeople}
\usetikzlibrary{
fit, hobby, backgrounds, shadings, shapes.arrows, shadows, patterns.meta,
shapes.geometric, positioning, arrows.meta, shapes.multipart, shapes.symbols, fit, calc}

\usepackage{pgf-pie}
\usepackage{pgfplots}
\pgfplotsset{compat=1.8}
\usepackage[group-separator={,}]{siunitx}
\usepackage{listings}
\usepackage{csquotes}
\usepackage{multirow}
\usepackage{pgf-umlsd}
\usepackage{xcolor}

\def\tikzmark#1{\tikz[remember picture,overlay]\node[yshift=2pt](#1){};}
\definecolor{codegreen}{rgb}{0,0.6,0}
\definecolor{codegray}{rgb}{0.5,0.5,0.5}
\definecolor{codepurple}{rgb}{0.58,0,0.82}

\usepackage[]{hyperref}

\begin{document}

\date{}

\newcommand{\sys}{\textsc{MatR}\xspace}
\newcommand{\relmap}{RelocationTable\xspace}

\newcommand{\numworkloads}{three\xspace}
\newcommand{\numvignettes}{three\xspace}
\newcommand{\latencySpeedupGeo}{2.19\xspace}
\newcommand{\loadingSpeedupGeo}{2.96\xspace}
\newcommand{\latencyspeeduplow}{1.13\xspace}
\newcommand{\latencyspeeduphigh}{7.03\xspace}
\newcommand{\microbenchmarkspeeduphigh}{15.3\xspace}

\setlist{leftmargin=*,itemsep=0pt,parsep=0pt,topsep=3pt}

\crefname{algocf}{algorithm}{algorithms}
\Crefname{algocf}{Algorithm}{Algorithms}
\crefformat{section}{\S#2#1#3}
\Crefformat{section}{Section~#2#1#3}
\crefformat{subsection}{\S#2#1#3}
\Crefformat{subsection}{Section~#2#1#3}
\crefformat{subsubsection}{\S#2#1#3}
\Crefformat{subsubsection}{Section~#2#1#3}
\crefformat{figure}{Figure~#2#1#3}
\Crefformat{figure}{Figure~#2#1#3}
\crefformat{table}{Table~#2#1#3}
\Crefformat{table}{Table~#2#1#3}

\newenvironment{smenumerate}%
  {\begin{enumerate}[itemsep=-0pt, parsep=0pt, topsep=0pt, leftmargin=2pc]}
  {\end{enumerate}}

\newif\ifcomments
\commentstrue

\newcommand{\arq}[1]{\ifcomments\textcolor{orange}{[[AQ: #1]]}}
\definecolor{FZColor}{HTML}{000088}
\newcommand{\fz}[1]{\ifcomments\textcolor{FZColor}{[[FZ: #1]]}}
\definecolor{MWColor}{HTML}{000077}
\newcommand{\mw}[1]{\ifcomments\textcolor{MWColor}{[[MW: #1]]}}

\definecolor{TodoColor}{HTML}{FF5733}
\newcommand{\todo}[1]{\ifcomments\textcolor{TodoColor}{[[Todo: #1]]}}



\title{Symbol Resolution MatRs: Make it Fast and Observable with Stable Linking}

\author{Farid Zakaria}
\email{fmzakari@ucsc.edu}
\affiliation{
    \institution{University of California Santa Cruz}
    \city{Santa Cruz}
    \country{USA}
}
\author{Andrew Quinn}
\email{aquinn1@ucsc.edu}
\affiliation{
    \institution{University of California Santa Cruz}
    \city{Santa Cruz}
    \country{USA}
}
\author{Thomas R. W. Scogland}
\email{scogland1@llnl.gov}
\affiliation{
    \institution{Lawrence Livermore National Laboratory}
    \city{Livermore}
    \country{USA}
}


\begin{abstract}

Dynamic linking is the standard mechanism for using external dependencies since it enables code reuse, streamlines software updates, and reduces disk/network use.  Dynamic linking waits until runtime to calculate an application's relocation mapping, i.e., the mapping between each externally referenced symbol in the application to the dependency that provides the symbol.  Unfortunately, it comes with two downsides.  First, dynamic linking limits the performance of current systems since it can take seconds to calculate a relocation mapping for a large program.  Second, dynamic linking limits the dependency management of applications since it prevents a developer from accurately observing a relocation mapping except at runtime. 

This paper makes the key insight that the benefits conventionally attributed to dynamic linking---code reuse, streamlined software updates, and reduced disk/network use---are actually benefits of shared libraries.  Thus, we present \emph{stable linking}, a new mechanism for using dependencies that uses shared libraries to retain their benefits but eliminates the downsides of dynamic linking.  Stable linking separates a system's state into management times; when the system can be modified, and epochs when it cannot.  Stable linking calculates each application's relocation mapping at the beginning of each epoch, allows developers to inspect the relocation mapping during the epoch, and reuses the mapping for subsequent executions in the epoch.  We design and build \sys{}, the first stable linker.  We use \sys{} in \numworkloads{} workloads and show that it improves upon dynamic linking performance by a factor of \latencySpeedupGeo on average.  Additionally, we use the system in \numvignettes vignettes, or case-studies, that illustrate the system's improvements to dependency management.

\end{abstract}

\maketitle
\pagestyle{plain}

\section{Introduction}

Dynamic linking is the de facto mechanism for using external dependencies in today's software systems including Windows, MacOS, and most Linux distributions.  Dynamic linking is a mechanism for symbol resolution; for each external symbol (i.e., function or variable) that an application references, the linker identifies the location of the symbol in an external dependency (i.e., software library) and updates the application's memory accordingly.  This paper refers to each binding from an application's external reference to a value in a dependency as a \emph{symbol relocation} and refers to the set of symbol relocations for an application as a \emph{relocation mapping}.  Dynamic linking differentiates from other symbol resolution approaches in that it waits until runtime to resolve symbols, which facilitates code reuse, streamlines software updates, and reduces disk and network use~\cite{Agrawal:asplos2015,Bartell:oopsla2020}.  

Despite its widespread use, dynamic linking limits the performance and dependency management of applications. On the performance side, dynamic linking is surprisingly expensive; Becker et al. show that dynamic linking accounts for 74\% of the total runtime for short-lived programs~\cite{becker2018measuring}, and our analysis shows that large applications spend most of their startup time on dynamic linking (\cref{sec:motiv:dynamic-linking}).  Expensive startup is especially problematic considering its aggregate cost; systems can execute hundreds of thousands of processes each day~\cite{Wang:icse21}, thereby repeatedly paying the cost of dynamic linking.

While existing techniques can accelerate dynamic linking, they do not fully resolve dynamic linking performance problems, forgo dynamic linking's benefits, or are incompatible with standard system features.  For example, some systems compile external dependencies into an application using static linking (e.g., Slinky~\cite{Collberg:atc2005}), which resolves dynamic linking performance issues, but also eschews its benefits.  Other proposals optimize dynamic linking (e.g., direct binding~\cite{direct-binding:online}, GNU hash tables (\verb|DT_GNU_HASH|)~\cite{dt-gnu-hash:online}), but provide little performance benefit.  Finally, some systems  (e.g., Prelink~\cite{prelink:online, Jung:emsoft2007}, Software Multiplexing~\cite{Dietz:oopsla2018}) propose new linking strategies that resolve performance issues, but are incompatible with system features like Address Space Layout Randomization (ASLR)~\cite{Bhatkar:uss03} and binary-based software distribution. 

On the dependency management side, managing an application's dependencies is notoriously error-prone~\cite{van1994expert}.  The crux of the dependency management problem is that administrators lack the tools necessary to manage their system's dependencies.  There are utilities that illuminate the relocation mapping that a dynamic linker would create given a system's current state (e.g., readelf~\cite{readelf111:online}, libelf~\cite{LibElfFr14:online}, llvm-ifs~\cite{llvmifss19:online}). However, these tools are fundamentally inaccurate since they can only illuminate an application's \emph{current} relocation mapping, but the  application's relocation mapping may change during a \emph{subsequent} execution since dynamic linking occurs at runtime.  Consequently, dependency management relies on best practices~\cite{fedora-abi:online} that fall short and have led to application failures~\cite{1457749–15:online, 1081252l15:online, 2136800–39:online} and security exploits~\cite{NVDCVE2056:online, CVERecor33:online, CVERecor18:online}. 

In this paper, we make the key insight that benefits that are conventionally attributed to dynamic linking---code reuse, streamlined security updates, and reduced disk/network use---are \emph{actually} benefits of using shared libraries. Since the aforementioned performance and dependency management limitations arise from dynamic linking---not from shared libraries---we argue that it is possible to return the benefits of shared libraries without facing the costs of dynamic linking.

1cBased upon this insight, we propose a new technique for dependency resolution called \emph{stable linking}.  Stable linking retains the use of shared libraries and thus retains their benefits.  Stable linking introduces a new approach to resolving symbols; rather than resolve symbols at compile time or runtime, stable linking resolves symbols when the system is in a special state called \emph{management time}.   Management times occur after compilation but before runtime; a user initiates a management time before modifying a set of dependencies on their system and terminates it once they finish with the updates.  

At the conclusion of a management time, stable linking calculates the resolution mapping for all applications based upon the new dependencies in the system.  When a user executes a program during an epoch (i.e., when the system is not in management time) stable linking loads the relocation mapping calculated at the end of the most recent management time, exploiting the fact that the relocation mapping cannot have changed since the system's dependencies have not changed.  Stable linking improves performance compared to dynamic linking since loading a resolution mapping is faster than calculating one.  Stable linking improves dependency management compared to dynamic linking by exposing each application's resolution mapping to users.

Stable linking requires that system dependencies remain unmodified outside of management time, else the approach would fail to load applications correctly.  Conveniently, system administration best practices (e.g., performing updates at scheduled times~\cite{limoncelli2007practice, BestPrac54:online}), package managers (e.g., Apt~\cite{apt:online}, Homebrew~\cite{Mcquaid:homebrew:online}, etc.), and emerging build systems (e.g., Nix~\cite{Dolstra:icfp2008, Dolstra:LISA2004, NixNixOS90:online}, Spack~\cite{Gamblin:hpc2015, SpackSpa8:online}, and Guix~\cite{GNUGuixt90:online}) limit system updates to well-defined times.  Thus, integrating stable linking into most systems is straightforward.

We design and implement a stable linker called \sys{} (\underline{Mat}erialized \underline{R}elocations). \sys{} tracks the applications and shared libraries on a user's system, allowing a user to modify them only during management time.  At the conclusion of each management time, \sys{} materializes the relocation mapping for each application by observing the relocation mapping produced by dynamic linking.  During executions that occur in the subsequent epoch, the system reuses the observed mapping.  
The system encodes each relocation mapping as a table that supports ASLR by tracking data offsets instead of memory addresses.  The system accelerates application startup by loading a program using these tables.  The system improves dependency management by exposing the table in three digestible formats: JSON, CSV, and as a queryable SQL table.

We evaluated \sys{}'s performance on \numworkloads{} real-world benchmarks: Clang, LibreOffice, and Lawrence Livermore National Laboratory's Pynamic benchmark~\cite{lee2007pynamic}.  We find that it accelerates application latency compared to dynamic linking by a factor between \latencyspeeduplow{} and \latencyspeeduphigh{}, with a geometric mean of \latencySpeedupGeo{}.  Moreover, we construct a microbenchmark and find that \sys{} accelerates application startup by a factor of \microbenchmarkspeeduphigh{} for applications that include \num{1000000} symbols. 

We evaluated \sys{}'s improvement in dependency management by using the system in \numvignettes{} \emph{vignettes}, or use-cases.  In the first vignette, a administrator uses \sys{}'s SQL interface to ensure ABI compatibility across two versions of a library.  In the second vignette, a data-center administrator uses \sys{}'s SQL interface to identify if any of the applications deployed in their data-center were susceptible to a zero-day CVE.  In the final vignette, a developer uses \sys{} to diagnose bugs more efficiently by applying memory debugging tools selectively on a buggy software library.

Finally, we identify two software practices that are used today because of inefficient symbol resolution; \sys{} can improve upon these practices since it accelerates symbol resolution.  First, today's application developers put significant effort into reducing the number of external symbols that their programs use to reduce symbol resolution cost~\cite{drepper2006write, Controls12:online}.  Since \sys{} can efficiently load applications that use millions of symbols, it alleviates the need for such efforts.  Second, current systems use lazy binding to accelerate symbol resolution at the cost of additional runtime overhead throughout a program execution.  Whereas prior approaches to accelerate lazy binding require custom architecture~\cite{Agrawal:asplos2015} or new compilers~\cite{Ren:osdi2022}, \sys{} enables a simple strategy: disable it!

In summary, our contributions are: 
\begin{itemize}
\item Stable linking; a new technique that retains the benefits of shared libraries but improves the performance and dependency management of dynamic linking. 
\item The design and implementation of \sys{}~\footnote{available at \url{https://github.com/fzakaria/musllibc/tree/management-time}}; a stable linker that materializes symbol relocation during each management time, loads them to accelerate startup, and exposes them to improve dependency management.
\item An evaluation of \sys{} on \numworkloads workloads showing that it improves application startup and \numvignettes{} vignettes showing that it improves dependency management.
\end{itemize}
\section{Motivation}
\label{sec:motiv}

This section motivates \sys{} by describing the two key limitations of dynamic linking: its high startup cost and its poor dependency management. Namely, we discuss background on dynamic linking (\cref{sec:motiv:dynamic-linking}), its performance limitations (\cref{sec:motiv:perf}), and its dependency management limitations (\cref{sec:motiv:management}). 

\subsection{Dynamic Linking Background}
\label{sec:motiv:dynamic-linking}

Dynamic linking is the standard mechanism for using external dependencies in a software system;  Windows, MacOS, and most Linux distributions use dynamic linking.  Dynamic linking is a form of late binding in that it delays an application's symbol resolution until runtime.  

The dynamic linking process requires that specific actions are taken during compile time by a linker~\cite{specification1993tool}, and at runtime by a dynamic linker~\cite{matz:sysv2013}.  When compiling a shared object, such as an application or library, a linker identifies the external symbols that the shared object references, including functions and variables.  The linker adds relocation instructions, which specify the external symbols that the shared object requires, to special sections in the shared object binary.  Additionally, the linker embeds a hash table that includes all of the symbols contained within the shared object. Note that while the relocation instructions include the external symbol name, they do not specify the library in which it will be found.

When running an application, the dynamic linker processes the relocation instructions.  It loads all of the application's external dependent libraries into memory. For each relocation instruction, the dynamic linker finds the location of the referenced symbol by iterating through its external dependencies in the order specified by the application and the user's execution environment, using the dependency's hash table to find the first library that contains the symbol, and updating the application's memory accordingly.  

\subsection{Performance Limitations}
\label{sec:motiv:perf}

Dynamic linking is surprisingly expensive, especially since it degrades system efficiency every time a user starts a process.  Moreover, dynamic linking scales extremely poorly to large applications---it takes dynamic linking more than 10 seconds to startup some of today's large applications.  Below, we benchmark dynamic linking to analyze its cost and then describe the limitations of prior approaches that aim to accelerate it.

\subsubsection{Benchmarking Dynamic Linking}
\label{sec:motiv:benchmark-dynamic}
We create a microbenchmark to identify the cost of Dynamic Linking.  The microbenchmark takes two parameters: a number of shared library dependencies, $n$, and a number of external functions, $f$.  It creates $n$ shared libraries and defines $f/n$ functions in each library; each function includes a single call to \texttt{printf}, while the benchmark's \texttt{main} function calls each of the defined functions.  We vary the number of shared objects from 1 to \num{10000} by powers of 10, and the number of total functions from 1 to \num{1000000} by powers of 10.  We execute the program on an AMD EPYC 7452
32-Core Processor running Linux 6.2.0 with a Micron SATA 6 Gbps SSD when using \texttt{musl}~\cite{musllibc83:online} and measure the execution time. 

\begin{figure}
\centering
\includegraphics[width=\columnwidth]{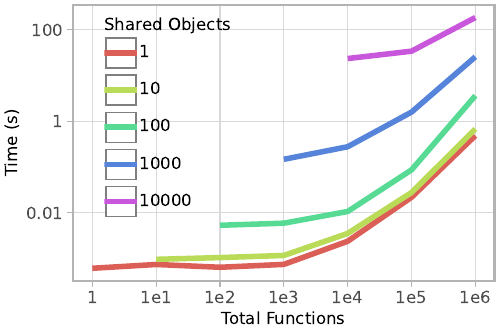}
\caption{Startup time for microbenchmark with varied shared object dependencies and varied number of external functions.}
\label{fig:unpatched-histogram}
\Description[Graph of startup time for applications with varying number of symbol relocations across multiple shared objects.]{A histogram demonstrating the startup cost for applications with increasing number of functions across multiple shared objects.}
\end{figure}

\Cref{fig:unpatched-histogram} shows the results.  Execution time is negligible for binaries that use fewer than 10 dependencies and fewer than \num{10000} external functions.  However, execution time scales poorly. It grows superlinearly with the number of external functions (e.g., increasing the number of external functions by a factor of 10 increases startup time by more than a factor of 10), and roughly linearly with the number of dependent shared libraries (e.g., increasing the number of shared objects by a factor of 10 increases the startup time by a factor of 10).  Thus, dynamic linking imposes non-negligible cost on applications with at least \num{10000} symbols, especially considering that some systems use dynamic linking on each of their more than \num{100000} processes each day~\cite{Wang:icse21}. Moreover, dynamic linking startup time reaches more than 10 seconds for applications that use thousands of dependencies or millions of external functions.

\begin{table}
    \centering
    \begin{tabular}{|l|l|l|}
    \hline
     Application & Relocations & Dependencies \\ 
     \hline\hline
     \verb|ls| \small{(9.3)} & 1649 & 4 \\ 
     \verb|ruby|  \small{(3.1.5)}& 2390 & 3 \\  
     \verb|python|  \small{(3.11.8)}& 14192 & 2 \\
     \verb|clang|  \small{(16.0.6)}& 44505 & 10 \\
     \verb|LibreOffice| \small{(7.6)}& 165985 & 138 \\
     \verb|pynamic| \small{(bigexe)} & 1807246 & 911\\\hline
    \end{tabular}
    \caption{Symbol relocations by application.}
    \Description[Symbol relocations by application.]{The number of symbol relocations that can be found across a sample set of applications.}
    \label{fig:symbol-relocation-by-app}
\end{table}

We contextualize the microbenchmark results by analyzing the number of relocations and dependencies that popular applications reference; \cref{fig:symbol-relocation-by-app} shows the results.  We observe the following by using \cref{fig:symbol-relocation-by-app} and \cref{fig:unpatched-histogram}.  First, there are some applications that experience negligible dynamic linking overhead (e.g., \texttt{ls} and \texttt{ruby}). Second, there are commonly used applications~\cite{Wang:icse21}, such as compilers (\texttt{clang}) and GUI applications (e.g., \texttt{LibreOffice}), that experience meaningful dynamic linking overhead, especially considering their frequent use.  Finally, some applications (e.g., \texttt{pynamic}) use enough relocations and external dependencies to experience more than 30 seconds of dynamic linking overhead.

\begin{figure}
\centering
\begin{tikzpicture}
    \pie[radius=1.75,
    text = legend,
    explode = {0, 0.1, 0}, 
    align=left
    ]{12.31/Loading Libraries,
    78.46/Symbol Relocation,
    9.23/Other}
\end{tikzpicture}
\caption{Time spent at startup by OpenOffice performing symbol relocations profiled with perf.}
\label{fig:open-office-pie}
\Description[Pie chart of time spent starting up for openoffice.]{Pie chart of time spent starting up for openoffice.}
\end{figure}
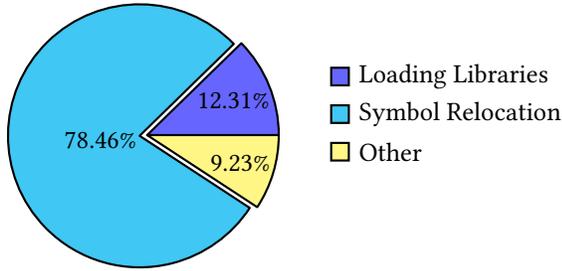

We analyze the cause of long startup cost by profiling \texttt{LibreOffice}'s startup.  We measure the time it takes to execute \texttt{LibreOffice}'s \texttt{--help} command line operation and use \texttt{perf}~\cite{de2010new} to profile the execution.  Figure~\ref{fig:open-office-pie} categorizes how \texttt{LibreOffice} spends its time.  Nearly 80\% of \texttt{LibreOffice} startup time is spent processing symbol relocations, i.e., identifying the dependent library that includes each symbol.

We conclude that processing symbol relocations is the dominant cost in dynamic linking and the cause of its poor scalability.  Drepper et al.~\cite{drepper2006write} analyzed the cost of external symbol relocation and found that it has super-linear scalability, conforming to \cref{fig:unpatched-histogram}.  They formalize external symbol relocation cost as $\mathcal{O}(nr\log{}s)$ where~$r$ is the number of external symbol relocations, $n$ is the number of shared objects (i.e., the executable and its dependencies), and $s$ is the number of symbols across all shared libraries and the executable. 

\subsubsection{Existing Mitigation Strategies}
\label{sec:motiv:existing-mitigation}

The community has developed several mitigation strategies to cope with dynamic linking's cost.  Unfortunately, existing strategies fail to resolve the performance problems, forgo shared library benefits, or are incompatible with current systems.

\paragraph{Lazy Binding}  Most dynamic linkers (e.g., glibc) default to using lazy binding to reduce overhead.
Lazy binding only resolves the external symbols referenced by the current execution by deferring symbol resolution until first use.  It reduces the cost of symbol relocation, but requires additional data structures that impose their own runtime overhead.  We describe and evaluate lazy binding in additional detail in the discussion (\cref{sec:discussion}).

\paragraph{Prelink}
Prelink reduces dynamic linking startup time by resolving relocations before execution~\cite{prelink:online}.  The tool scans an executable and its shared libraries, assigns fixed virtual memory addresses to each library, computes the relocation values based on the fixed memory layout, and updates the executable and libraries with the relocation values.  Prelink accelerates startup, but is incompatible with current systems since its fixed virtual memory addresses do not support address space layout randomization~\cite{Bhatkar:uss03}. 


\paragraph{Optimizing Symbol Resolution}

Many techniques accelerate dynamic linking by optimizing its symbol resolution.  While these approaches alleviate some of the dynamic linking cost, they do not eliminate it.  One approach optimizes dynamic linking's hash tables to accelerate the process of looking up a symbol.  Developers have optimized the hash table's data layout, hash function, and even integrated a bloom filter that accelerates identifying the dependency that includes each symbol~\cite{PATCHDTG46:online, Bloom:cacm1970, JakubJel45:online}. Other approaches~\cite{direct-binding:online} encode the desired dependent library name alongside the symbol so a dynamic linker only searches for the symbol in the right dependency. Unfortunately, symbol relocation remains expensive, even when narrowed to a single library.

\paragraph{Static Linking}
Finally, some systems forgo dynamic linking altogether by statically linking external dependencies into an application.  Unfortunately, static linking forgoes the benefits of shared libraries causing cumbersome updates, reduced code reuse, and more disk/network use~\cite{StaticLi94:online}.  

\subsection{Dependency Management Limitations}
\label{sec:motiv:management}

Dynamic linking camouflages an application's relocation mapping, thereby complicating the process of auditing and debugging an application's dependencies.  One potential solution is to execute an application and observe the relocations made by the dynamic linker.  There are two issues with this approach. First, lazy binding means that there may be relocations that the program did not perform during the traced execution.  Second, since dynamic linking occurs at runtime, an application's resolution mapping can change between any two executions.  Thus, a resolution mapping from one execution may not accurately reflect the resolution mapping in the next.

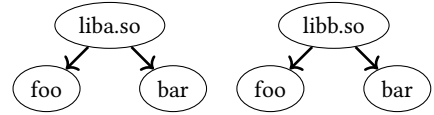
\begin{figure}
      \begin{tikzpicture}
        \small
    \tikzstyle{oval} = [draw, shape=ellipse, rounded corners, minimum height = 6mm]
    \tikzstyle{line} = [line width=1pt]

    \node[oval] (a) {liba.so};
    \node[oval, right=15mm of a] (b) {libb.so};

    \node[oval, below left=4mm and 0mm of a] (fooa) {foo};
    \node[oval, below right=4mm and 0mm of a] (bara) {bar};
    \draw[line, ->] (a) to (fooa);
    \draw[line, ->] (a) to (bara);

    \node[oval, below left=4mm and 0mm of b] (foob) {foo};
    \node[oval, below right=4mm and 0mm of b] (barb) {bar};
    \draw[line, ->] (b) to (foob);
    \draw[line, ->] (b) to (barb);
\end{tikzpicture}
    \caption{Dynamic linking cannot bind $foo$ from $liba.so$ and $bar$ from $libb.so$}
    \Description[Paradoxical setup impossible with today's dynamic-loader]{Paradoxical setup impossible with today's dynamic-linker.}
    \label{fig:paradox}
\end{figure}

In addition to limiting visibility into the resolution mapping, dynamic linking also limits a user's ability to update their relocations.  Dynamic linking uses a single ordered list to decide which dependency to use when binding a symbol to a relocation, which limits a user's flexibilty.  For example, suppose a user wishes to use the symbol \texttt{foo} from the library \texttt{liba.so} and the symbol \texttt{bar} from the library \texttt{libb.so}.  If both libraries export both symbols, the user cannot use their desired setup with dynamic linking (see \cref{fig:paradox}).  The same limitation prevents a user from performing partial resolutions---e.g., a user cannot setup their application so that calls to \texttt{malloc} made from a shared library use a custom implementation in \texttt{libmalloc.so}, while all other calls to \texttt{malloc} use the default implementation in \texttt{libc}.

\subsection{Key Takeaways} In summary: 
\begin{itemize}
    \item Dynamic linking imposes high startup cost, especially considering the aggregate cost of process creation.
    \item Dynamic linking obscures relocation mappings.
    \item Proposed solutions to these problems fail to resolve them, forgo the benefits of dynamic linking, or are incompatible with system features.
\end{itemize}
\section{Stable Linking}
\label{sec:stable}
\label{sec:stability}

\begin{figure}
    \begin{tikzpicture} [
      level distance = 10mm,
      edge from parent path={
        (\tikzparentnode) |-
        ($(\tikzparentnode)!0.5!(\tikzchildnode)$) -|
        (\tikzchildnode)}]
        \small
    \tikzstyle{boxy} = [shape=rectangle, rounded corners, draw, fill=white, minimum height=3ex, align=center, anchor=south]
    \tikzstyle{box}  = [shape=rectangle, draw, fill=white, minimum height=2cm, align=left]
    \tikzstyle{mgmt} = [fill=blue!20, shape=rectangle, minimum width=18mm, minimum height=15mm, 
    label={[anchor=center, align=center]center:Management\\Time}]
     \tikzstyle{epoch} = [fill=white, shape=rectangle, minimum height=15mm,     label={[anchor=south east]south east:Epoch}]

    \tikzstyle{app} = [boxy, label={[anchor=south west]south west:\footnotesize App}, minimum width=10mm, minimum height=9mm]
    
    \tikzstyle{line} = [line width=1.25pt]
    \tikzstyle{link} = [line, <->]

    \node (time) {Time};
    \node[right=75mm of time] (end) {};
    \draw[line, ->] (time) to (end);

    \node[above right=1mm and 0mm of time.east, mgmt, anchor=south west] (mgmt1){};

    \node[right=0mm of mgmt1.east, epoch, anchor=west, minimum width=15mm] (e1){};

    \node[right=0mm of e1.east, mgmt, anchor=west] (mgmt2){};

    \node[right=0mm of mgmt2.east, epoch, anchor=west, minimum width=22mm] (e2) {};


    \node[app, right=1mm of mgmt1.north east, anchor=north west] (a1) {\footnotesize};
    \node[boxy, below left=1mm of a1.north east, anchor=north east] {\footnotesize RM};

    \node[app, below right=1.5mm of a1.north west, anchor=north west] (a2) {\footnotesize};
    \node[boxy, below left=1mm of a2.north east, anchor=north east] {\footnotesize RM};


    \node[app, right=1mm of mgmt2.north east, anchor=north west] (a4) {\footnotesize};
    \node[boxy, below left=1mm of a4.north east, anchor=north east] {\footnotesize RM};

    \node[app, below right=1.5mm of a4.north west, anchor=north west] (a5) {\footnotesize};
    \node[boxy, below left=1mm of a5.north east, anchor=north east] {\footnotesize RM};


\end{tikzpicture}
  \caption{System state over time with stable linking.  RM stands for Relocation Mapping; App stands for application.
  \label{fig:management}}
\Description[TBD]{TBD}
\end{figure}
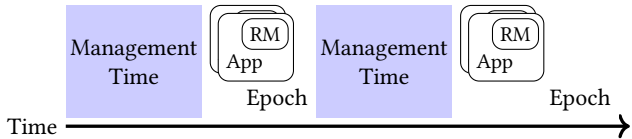

Our key insight is that conventional wisdom confounds dynamic linking and shared object libraries.  Consequently, the advantages of shared libraries---code reuse, streamlined security updates, and reduced disk/network use---are frequently attributed to dynamic linking~\cite{StaticLi44:online}, while the disadvantages of dynamic linking---limited performance and poor dependency management---are frequently attributed to shared libraries~\cite{StaticLi20:online}.  However, the two ideas are distinct: it is possible to retain shared libraries but replace dynamic linking, thus retaining the benefits of shared libraries without the downsides of dynamic linking. 

We propose stable linking as a new technique for resolving external dependencies in applications.  Stable linking uses shared libraries to organize dependencies---thereby retaining their benefits---and provides a solution to the performance and dependency management problems of dynamic linking.  Stable linking differentiates a system's state into \emph{management times}, which are periods of time when the dependencies and applications on a system can change, and \emph{epochs}, which are periods between management times.  At the start of each epoch, stable linking generates the relocation mapping for each application, which is guaranteed to remain constant throughout the epoch.  \Cref{fig:management} depicts how a system's state changes with stable linking over time.

Stable linking resolves dynamic linking's performance and dependency management issues by exploiting the fact that an application's relocation mapping is constant throughout an epoch since the system's dependencies do not change during the epoch.  Namely, stable linking accelerates dynamic linking performance by loading the constant relocation mapping without needing to perform expensive symbol resolution (see \cref{sec:motiv:perf}).  Stable linking improves dependency management by exposing accurate relocation mappings, since the mappings cannot change until the next management time.

Below, we describe how static linking performs each of the core operations and also outline how to extend existing systems to use stable linking.

\paragraph{Management Time} Stable linking allows a user to update an application's dependencies without requiring the application to be rebuilt. Management times occur after compilation, but before runtime.  A user creates a new management time by calling \texttt{begin\_mgmt} and ends the current management time by calling \texttt{end\_mgmt}.  In between these calls, a user modifies the dependencies and applications on their system.  The user informs the stable linker of a system modification to an object on their system by calling \texttt{update\_obj}.  When management time ends, but before the next epoch begins, a stable linker determines the relocation mapping of each of the applications on the system. 

\paragraph{Execution} When a user executes an application managed by a stable linker, the linker resolves application dependencies by loading the precomputed symbol resolution mapping from the most recent management time.  Loading a precomputed symbol resolution mapping is significantly faster than performing traditional symbol resolution, so stable linking drastically accelerates the process of symbol resolution during runtime compared to dynamic linking.  Note that stable linking cannot accelerate any symbol resolution that an application performs using tools such as \texttt{dlopen} since it may not have observed the resolutions at the end of the previous management time.

\paragraph{Relocation Mapping} Stable linking exposes the relocation mapping to end-users, providing a better platform for updating, auditing, and debugging an application's symbol resolution. Stable linking allows visibility into the mapping that is not possible with today's tools since the approach ensures that the symbol resolution mapping accurately reflects an application's symbol resolutions.  Moreover, stable linking allows advanced users to perform fine-grained modifications to the relocation mapping (e.g., updating a single relocation) which is not possible today. 

\paragraph{Integration} Stable linking requires that system applications and dependencies remain unmodified outside of a management time.  Conveniently, common approaches to managing system dependencies lend themselves to enforcing management time, so porting an existing system to use stable linking is straightforward.  For example, when a user updates their system via a package manager (e.g., Homebrew, Apptitude, etc.), the manager would execute \texttt{begin\_mgmt} before downloading any packages, call \texttt{update\_obj} for each downloaded dependency and application, and execute \texttt{end\_mgmt} after finishing its update.

\section{Design \& Implementation}
\label{sec:mater}

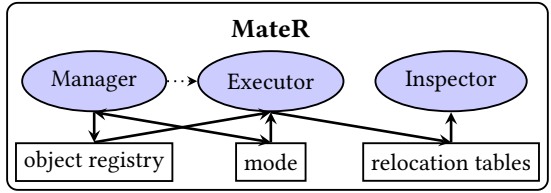
\begin{figure}
    \begin{tikzpicture} [
      node distance = 2mm,
      line width=.25mm]
    
    \tikzstyle{executor} = [draw, shape=ellipse, rounded corners, minimum height = 8mm]
    \tikzstyle{external} = [executor, fill=blue!20]

    \tikzstyle{state} = [draw, fill=white, shape=rectangle, minimum height = 5mm, align=left]

    \tikzstyle{data} = [->, line width=1pt, >=stealth, align=center]
    \tikzstyle{control} = [->, >=stealth, align=center, dotted]

    \small

    \node[external] (manage) {Manager};
    \node[external, right=4mm of manage] (exec) {Executor};
    \node[external, right=4mm of exec] (inspect) {Inspector};


    \node[state, below=4mm of manage] (registry) {object registry};
    \node[state, below=4mm of exec] (mode) {mode};
    \node[state, below=4mm of inspect] (tables) {relocation tables};

    \draw[data, <->] (manage.south) to (mode.north);
    \draw[data] (manage.south) to (registry.north);
    \draw[control] (manage.east) to (exec.west);

    \draw[data]  (mode.north) to (exec.south);
    \draw[data] (registry.north) to (exec.south);
    \draw[data] (exec.south) to (tables.north);

    \draw[data] (tables.north) to (inspect.south);

    \node[above=.5mm of exec] (label) {\textbf{\normalsize MateR}};




   \node[draw, shape=rectangle, rounded corners, fit=(label)(manage) (exec) (inspect)(mode)(tables)(registry)] (sBox) {};

\end{tikzpicture}
    \caption{\sys{}'s design. Blue ovals show active components of the system, white boxes show the system state.  Arrows indicate data flow, while dashed arrows indicate control flow.
    \label{fig:design}}
    \Description[]{}
\end{figure}

We design and implement \sys{}, the first stable linker.  \sys{} tracks the shared objects in the system, consisting of both dependent shared libraries and applications.  At the end of each management time, \sys{} materializes the relocations for each application on a system at the end of each management time, which it stores as a \emph{relocation table}.  Since stable linking ensures that an application's dependencies, and hence relocation mapping, cannot change during an epoch, \sys{} uses the relocation table during the subsequent epoch.  During execution, the system loads the table into memory and iterates through each of its entries to perform relocations.  Crucially, \sys{} supports ASLR's randomized memory locations~\cite{Bhatkar:uss03} by encoding each symbol relocation using offsets instead of memory addresses. Overall, \sys{}'s resolution accelerates application startup since iterating the relocation table is much more efficient than the symbol resolution of a traditional dynamic linker. 

\sys{} provides dependency management by exposing its relocation tables to end-users in three data formats: CSV, JSON, and SQLite.  The relocation tables provide users with accurate inspection since the tables will not change during an epoch.  In addition to viewing the tables, \sys{} allows advanced users to edit relocation tables to perform fine-grained symbol resolution updates which are not possible with dynamic linking.  Later, we show how the enhanced functionality can prevent dependency issues, manage security updates, and improve application interposition (\cref{sec:eval:dependency}).

Our implementation of \sys{} uses the Nix build system~\cite{NixNixOS90:online} to provide stability and extends musl~\cite{musllibc83:online} to perform linking. \Cref{fig:design} provides a high-level design diagram.  The system consists of three core execution components (the Manager, Executor, and Inspector) and three core collections of state (the object registry, mode, and relocation tables).  Below, we describe the Manager (\cref{sec:design:management}), Executor (\cref{sec:design:runtime}), and Inspector (\cref{sec:design:management}).

\subsection{The Manager}
\label{sec:design:management}

The Manager performs two main tasks.  First, it tracks the applications and shared libraries on the system through the object registry.  Second, the Manager updates the mode from being either management time or epoch.  Users interact with the Manager through three interfaces.  When a user invokes \texttt{begin\_mgmt}, the Manager changes the mode to management time.  When a user invokes \texttt{update\_obj}, the Manager ensures that the mode is set to management time and then updates its object registry to include the updated version of the object.  Critically, the Manager prevents a user from updating the object registry when the mode is set as epoch.  Finally, when a user invokes \texttt{end\_mgmt}, the Manager changes the mode to epoch and invokes the Executor with a special flag, \texttt{materialize}, indicating that the Executor should create the relocation tables (see \cref{sec:design:runtime}). 

Our Manager implementation uses Nix~\cite{Dolstra:icfp2008, Dolstra:LISA2004}.  Nix is a store-based system originally designed to ensure that the process of building an application is reproducible and deterministic.  To update an application on the system, a Nix user creates a build recipe that specifies the steps required to build the application, including compilation and configuration.  Then, Nix follows the recipe, building the application and all of its applications' dependencies, tagging each binary (i.e., executable or shared object) with the cryptographic hash of its content, and placing them into a read-only directory called the Nix store.  Nix sets the application's dependencies to point to libraries in the store tagged with the build-time cryptographic hashes.

While originally designed to simplify software builds, we use Nix to provide the core functionality of \sys{}'s Manager.  The Nix store corresponds to \sys{}'s object registry.  A user begins management time (\texttt{begin\_mgmt}) when they update a build recipe and invoke Nix, which ends (\texttt{end\_mgmt}) when Nix finishes with the recipe.  \sys{} extends Nix in one key way: it invokes the \sys{} Executor with the \texttt{materialize} flag after Nix finishes building a recipe, but before returning to the user. 

\begin{figure}
    \centering
    \begin{minted}[
    xleftmargin=1em,
        frame=lines,
        fontsize=\small,
        linenos=true, 
        escapeinside=!!, 
        highlightlines={}
        ]{c++}
typedef struct {
  // How to process the relocation. 
  // From the object that requires the symbol
  int type;
  size_t addend;
  size_t offset;
  
  // Where the symbol is located.
  // From the object that provides the symbol.
  size_t st_value;
  size_t st_size;
  
  // UUID of the shared objects that require 
  // and provide the symbol.
  size_t requires_so_UUID; 
  size_t provides_so_UUID;
  
  // Inspector information
  char symbol_name[PATH_MAX];
  char requires_so_name[PATH_MAX];
  char provides_so_name[PATH_MAX];
} RelocationTableItem;
    \end{minted}
    \caption{The struct for each entry in a relocation table.}
    \label{fig:cached-reloc-info-struct}
    \Description[]{}
\end{figure}

\subsection{The Executor}
\label{sec:design:runtime}

The Executor is responsible for starting an execution of the applications within the object registry.  The Executor has three modes of operation depending upon the value of the mode and its invocation flags.  When the mode is management time, the Executor loads the requested application using traditional dynamic linking,
which ensures that an application's behavior is correct during updates.  When the Manager invokes the Executor with the \texttt{materialize} flag, the Executor materializes relocation tables for all entries in the object registry that were updated since the last epoch.  Finally, when the mode is set to epoch and the \texttt{materialize} flag is unset, the Executor uses the relocation table to load the requested application. 

\paragraph{Materialization}
The Executor creates relocation tables for all entries in the object registry that were updated since the last epoch when invoked with the \texttt{materialize} flag.  In essence, materialization stores the relocation mapping produced by an invocation of traditional dynamic linker.  Since the system's state cannot change during the epoch, \sys{} can safely reuse the relocation mapping produced by the materialization in subsequent executions of the application during the same epoch.  However, materialization cannot record exact memory addresses, since such an approach would be incompatible with ASLR.  Thus, \sys{} instead stores relative offsets that are accurate even when ASLR adjusts exact memory addresses. 

In particular, \sys{} iterates through the object registry and identifies all shared objects that are executable applications, as opposed to shared libraries.  It executes each application following traditional dynamic linking, with lazy binding disabled.  It assigns a universally unique identifier (UUID) to each loaded shared object, i.e., the executable and dependent libraries.  The Executor observes each symbol relocation and extracts necessary data into a \texttt{struct}; \Cref{fig:cached-reloc-info-struct} provides the definition.  Namely, \sys{} assigns \texttt{requires\_so\_uuid} to the UUID of the shared object that requires the symbol and \texttt{provides\_so\_uuid} to be the UUID of the shared object that provides the symbol.  It takes \texttt{offset}, \texttt{type}, and \texttt{addend} from the relocation instructions in the object that requires the symbol, and takes \texttt{st\_value} and \texttt{st\_size} from the symbol table of the object that provides the symbol.  Finally, it adds data useful for inspection: \texttt{symbol\_name}, \texttt{requires\_so\_name}, and \texttt{provides\_so\_name} are the name of the symbol, name of the shared object that requires the symbol, and name of the shared object that provides the symbol.

\paragraph{Execution}
When a user executes an application during an epoch, the Executor loads it using the application's relocation table.  It loads each of the shared objects into a random memory location, following ASLR, and saves the location into a hash table indexed by the unique UUIDs from materialization.  The Executor then performs each relocation by iterating through each entry in the relocation table, using the information stored in each entry and in the shared object hash table, and updating the application memory accordingly.  Notably, most of the Executor's memory operations are sequential and well suited for memory prefetching, whereas traditional dynamic linking involves multiple random accesses into sections of multiple shared objects.

\subsection{The Inspector}
\label{sec:design:visibility}

The Inspector exposes the relocation tables to enable better dependency management.  The Inspector exposes the relocation tables using the schema of the \texttt{RelocationTableItem} struct (\Cref{fig:cached-reloc-info-struct}) and provides three formats: JSON, CSV, and SQLite. The system also provides utilities to extract symbol table information from a shared object that is not contained in the object registry; e.g., it provides a table generation function, \texttt{ABI(library)}, which extracts the symbols exported by \texttt{library}, a shared object.  In addition, the system allows advanced users to modify the exposed relocation tables to update an individual symbol resolution, e.g., by modifying the \texttt{provides\_so\_UUID} field.  In our experience, the SQL interface is useful for dependency auditing symbols, while the flat file formats (CSV and JSON) are useful when performing fine-grained updates.
\section{Evaluation}
\label{sec:eval}

We evaluate \sys{} by answering the following questions:
\begin{itemize}[noitemsep]
    \item Does \sys{} accelerate application startup (\cref{sec:eval:startup})?
    \item Does \sys{} improve dependency management (\cref{sec:eval:dependency})? 
\end{itemize}


\begin{table}
    \centering
    \small
    \begin{tabular}{|l|l|l|l|}
    \hline
     Application & Workload & Relocations & Libraries \\ 
     \hline\hline
     \verb|clang|  \small{(16.0.6)}& AnghaBench &44505 & 10 \\
     \verb|LibreOffice| \small{(7.6)}& \texttt{--help} & 65985 & 138 \\
     \verb|pynamic| \small{(1.3.4)} & bigexe & 1807246 & 911\\
     \hline
    \end{tabular}
    \caption{Summary of the real-world benchmarks, including their workload, number of symbol relocations, and number of shared libraries.}
    \Description[Symbol relocations by application.]{The number of symbol relocations that can be found across a sample set of applications.}
    \label{tab:benchmarks}
\end{table}


\subsection{Experimental Setup} 
\label{sec:eval:setup}
We deployed and evaluated \sys{} on an AMD EPYC 7452 32-Core Processor running Linux 6.2.0 with a Micron SATA \SI{6}{Gbps} SSD.  We evaluate \sys{} using one microbenchmark (\cref{sec:eval:startup:microbenchmarks}) and \numworkloads{} real-world applications (\texttt{clang}, \texttt{LibreOffice}, and \texttt{Pynamic}).  For each application, we compare the application's performance when using \sys{} to the application when using musl~\cite{musllibc83:online}, which does not use lazy binding.  For each experiment, we perform 5 warmups, 10 trials, report mean values, and include P95 confidence intervals. \Cref{tab:benchmarks} summarizes the real-world benchmarks; we elaborate on them below.

\paragraph{Clang}  Code compilation is an extremely common task (e.g., Google executes millions of builds per day~\cite{Wang:icse21}), so even small savings in compilation latency can lead to large benefits.  We evaluate \texttt{Clang}~\cite{lattner2008llvm} when compiling C programs.  We create a workload using AnghaBench~\cite{9370322}, a collection of 1 million small C files from open-source codebases, that includes a sample of 100 random C files. 

\paragraph{LibreOffice}
\texttt{LibreOffice} is a suite of GUI-based office applications written in C++ and containing over 4 million lines of code across 10 thousand files.  We start the program with \texttt{--help} to evaluate the startup time that a user experiences when starting the GUI.

\paragraph{Pynamic}  
\texttt{Pynamic} is a benchmark suite developed by the Lawrence Livermore National Laboratory (LLNL)~\cite{lee2007pynamic} to model the behavior of loading real-world MPI software for scientific simulation.  We configure the benchmark to match the characteristics of real LLNL applications~\cite{pynamicc16:online}.

\subsection{\sys{} Startup Time}
\label{sec:eval:startup}
We evaluate \sys{}'s improvement in application startup time using microbenchmarks and real-world applications.  The microbenchmarks articulate \sys{}'s potential advantage, while the real-world applications show the system's advantage in real-world scenarios.  

\subsubsection{Microbenchmarks}
\label{sec:eval:startup:microbenchmarks}

We construct a microbenchmark to explore \sys{}'s latency improvements.  The benchmark accepts a number of shared objects, $n$, and number of functions, $f$, as input.  It creates $n$ new shared objects and defines $f$ functions in each shared object.  Each function performs a single call to \texttt{printf}, while the benchmark's \texttt{main} function executes each of the $n * f$ generated functions.  We generate test cases varying the number of shared objects and functions from 1 to \num{1000} by powers of 10.

\begin{figure}
\centering
\includegraphics[width=\columnwidth]{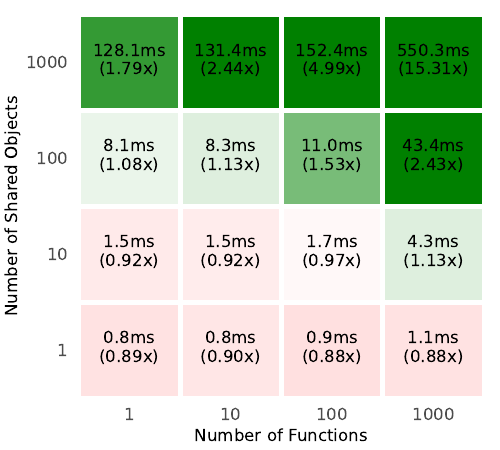}
\caption{Speedup when \sys has been applied.}
\label{fig:benchmark-heatmap}
\Description[Heatmap of speedup for different number of relocations.]{A matrix (heatmap) of various speedups applying the optimzation across two dimensions: number of functions per shared object and number of shared objects.}
\end{figure}

\Cref{fig:benchmark-heatmap} shows the execution latency of \sys{} and the system's speedup compared to \texttt{musl}.  \sys{} accelerates execution latency by between a factor of .8 and 15.3 compared to the baseline system.  \sys{}'s improvement increases as we increase the number of functions ($f$) and number of shared objects ($n$).  We note that \sys{}'s speedup is improved more when increasing $n$ than when increasing $f$.

\sys{} improves application startup for all scenarios where the  number of functions across all shared libraries ($n * f$) is greater than 1,000. However, when the number of total functions ($n * f$) is less than or equal to 1,000, \sys{} imposes a small slowdown (shown as a speedup less than 1.0).  \sys{} executes additional system calls to load its relocation table, which takes longer than \sys{}'s symbol resolution savings when there are few relocations to perform.  We note that the execution latencies for the scenarios where \sys{} slows execution are very small---in real terms, \sys{}'s slowdown is at most $120\mu$s. 


\begin{table}
\centering
\begin{tabular}{|l|l|l|l|l|}
\hline
Application & \texttt{musl} (s) & \sys (s) & Speedup\\ 
\hline\hline
\verb|Clang|       & $6.61$ {\footnotesize$\pm 0.05$} & $ 5.01$ {\footnotesize$\pm 0.07$}  & $1.32$ {\footnotesize$\pm 0.02$} \\
\verb|LibreOffice| & $0.58$ {\footnotesize$\pm 0.00$}& $0.51$ {\footnotesize$\pm 0.00$}  & $1.13$ {\footnotesize$\pm 0.01$} \\
\verb|Pynamic| & $31.80$ {\footnotesize$\pm 1.13$} & $4.524$ {\footnotesize$\pm 0.13$}  & $7.03$ {\footnotesize$\pm 0.20$} \\
\hline\hline
Geometric Mean & & & \latencySpeedupGeo{}\\
\hline
\end{tabular}
\caption{Real-world benchmark execution latency.     
\label{tab:real-benchmarks-only-summary}}
\end{table}

\begin{table}
\centering
\begin{tabular}{|l|l|l|l|}
\hline
 Application & \texttt{musl} (ms) & \sys (ms) & Speedup\\ 
 \hline\hline
 \verb|Clang| & $51$ {\footnotesize$\pm 0$} & $37$ {\footnotesize$\pm 0$} & $1.38$ {\footnotesize$\pm 0.01$} \\
 \verb|LibreOffice| & $151$ {\footnotesize$\pm 1$} & $71$ {\footnotesize$\pm 0$} & $2.14$ {\footnotesize$\pm 0.01$} \\
 \verb|Pynamic| & $30972$ {\footnotesize$\pm 1085$} & $3530$ {\footnotesize$\pm 159$} & $ 8.78$ {\footnotesize$\pm 0.40$} \\
 \hline\hline
 Geometric Mean & & & \loadingSpeedupGeo{}\\
 \hline
\end{tabular}
\caption{Application startup latency.\label{tab:real-benchmarks-donothing-summary}}
\end{table}

\subsubsection{Real-world Benchmarks} We use \texttt{musl} and \sys{} to execute \texttt{clang}, \texttt{LibreOffice}, and \texttt{Pynamic}; \Cref{tab:real-benchmarks-only-summary} shows the results across \texttt{musl} and \sys{}.  \sys{} accelerates application performance by a factor of between \latencyspeeduplow{} and \latencyspeeduphigh{}, with a geometric mean of \latencySpeedupGeo{}.  \sys{}'s improvements on \texttt{clang} and \texttt{LibreOffice} are useful considering that end-users frequently execute code compilation and office applications. \sys{} is especially impactful for \texttt{pynamic}, reducing latency from more than 30 seconds to less than 5 since  \texttt{Pynamic} performs close to 2 million symbol relocations.

Next, we isolate the time that each application spends during startup (i.e., before \texttt{main}) to illustrate \sys{}'s speedup.  We built an interposition library that implements \texttt{main} as a single line: \texttt{return EXIT\_SUCCESS;}.  We use \texttt{LD\_PRELOAD} to use the library when executing each application, short-circuiting the execution to ensure that it only executes application loading.  \Cref{tab:real-benchmarks-donothing-summary} shows the application startup results for \texttt{clang}, \texttt{LibreOffice}, and \texttt{pynamic}.  \sys{} provides a more pronounced improvement when we isolate startup.  \sys{} accelerates application startup by between a factor of 1.35 and 10.42, with a geometric mean of \loadingSpeedupGeo{}.  Once again, the \texttt{pynamic} benchmark is standout: \sys{} accelerates \texttt{pynamic} startup by more than an order of magnitude.

\subsection{Dependency Management}
\label{sec:eval:dependency}
\label{sec:vignettes}

We describe how \sys{} improves an administrator's ability to manage dependencies on their systems.  We articulate \numvignettes{} use-cases that illuminate the advantage of using \sys{} to observe and update a system's dependencies.  We present the use-cases as \emph{vignettes}, or user-stories, that describe qualitatively how an admin benefits from the system.  

\subsubsection{Vignette 1: ABI Compatibility}

A system administrator, Alice, wishes to update a shared library, \texttt{libfoo}, on their system.  Many of the applications on their system use \texttt{libfoo}, so Alice needs to ensure that none of them are incompatible with \texttt{libfoo}'s new Application Binary Interface (ABI), i.e., the set of symbols that it exports.  Unfortunately, dynamic linking complicates the process of determining ABI compatibility. Alice cannot accurately check her application for ABI compatibility since dynamic linking cannot accurately expose relocation mappings (\cref{sec:motiv:dynamic-linking}).  The net effect of these limitations is that Alice relies on the library maintainers to carefully scrutinize their libraries to prevent ABI changes~\cite{Howtoche20:online}, which often goes wrong~\cite{1457749–15:online, 2136800–39:online, 1081252l15:online}.

\begin{figure}
    \centering
    \begin{minted}[
    xleftmargin=1em,
        frame=lines,
        fontsize=\small,
        linenos=true, 
        escapeinside=!!, 
        highlightlines={}
        ]{sql}
SELECT RT.symbol_name, RT.object_name
FROM RelocationTable(App) as RT
LEFT JOIN ABI(libfoo) as ABI
ON RT.symbol_name = ABI.symbol_name
WHERE RT.provides_dso_name = "libfoo"
  AND ABI.symbol_name IS NULL
    \end{minted}
    \caption{ABI compatibility SQL query.}
    \label{fig:sql-join-abi}
    \Description[]{}
\end{figure}

\sys{} improves dependency management by allowing Alice to determine whether her applications are compatible with \texttt{libfoo}'s new ABI.  Alice writes a SQL query using \sys{}'s SQLite interface (\Cref{fig:sql-join-abi}).  Given an application argument (\texttt{App}), the query identifies \texttt{App}'s symbol resolutions against the old \texttt{libfoo}, where the resolution is against a symbol that is not contained in the new \texttt{libfoo}.  Executing the query over all applications allows Alice to check for ABI compatibility across her entire system.  The query joins the \relmap{} for \texttt{App} with \mintinline{sql}{ABI(libfoo)}, a table that encodes \texttt{libfoo}'s ABI that is provided by \sys{} (\cref{sec:design:management}).  Alice joins the table on \mintinline{sql}{symbol_name}; she uses a \texttt{LEFT JOIN} and filters on \texttt{ABI.symbol\_name IS NULL} to produce only those rows from \relmap{} that do not have a match.  Alice uses the query output to identify applications that she should rebuild and update when updating \texttt{libfoo}.

\subsubsection{Vignette 2: CVE Auditing}
\begin{figure}
    \centering
    \begin{minted}[
    xleftmargin=1em,
        frame=lines,
        fontsize=\small,
        linenos=true, 
        escapeinside=!!, 
        highlightlines={}
        ]{sql}
SELECT RM.object_name
FROM RelocationTable(App) as RT
WHERE RM.symbol_name = baz
  AND RM.dependency_name = "libbar"
    \end{minted}
    \caption{CVE Auditing SQL query}
    \label{fig:cve-query}
    \Description[]{}
\end{figure}

Bob, a data center administrator, receives notice of a privilege escalation CVE that affects a shared library, \texttt{libbar}, that some of the applications in the data center use.  The vulnerability lies in a single function, \texttt{baz}, within the larger shared object~\cite{FixCVE2022:online}.  Bob wishes to identify which applications were susceptible to the problem, quarantine the machines that execute those applications, and further investigate each quarantined machine's system log.  Unfortunately, Bob would not be able to identify the potentially compromised applications using dynamic linking since the approach cannot provide an accurate relocation mapping (\cref{sec:motiv:dynamic-linking}).  Consequently, Bob would either need to investigate all machines in the data-center or forgo the search altogether.

\sys{} provides functionality that drastically improves Bob's ability to audit the data-center for compromised systems.  Bob uses a centralized \sys{} stable linker on a management machine to prepare applications run on the data center, which ensures that the dependencies of each application are identical across machines.  On the management machine, Bob uses \sys{}'s SQL interface to select all applications that include a relocation to \texttt{libbar}'s \texttt{baz} function. \Cref{fig:cve-query} shows the query; Bob executes it over each application, \texttt{App}.  Bob uses the output of the query, together with a mapping of the machines that execute each application, to determine which machines to quarantine and investigate for compromise.

\subsubsection{Vignette 3: Fine-grained Interposition}

Charlie, a developer, faces a segmentation fault reported in the Apache Web Server~\cite{Apache45605}.  The issue only reliably arises in production, since it is a ``Heisenbug'' that depends upon a specific thread schedule~\cite{Musuvathi:osdi2008}, and the failure occurs much later than the fault, since it arises due to a memory corruption~\cite{Fonseca:dsn2010}.  Charlie would like to deploy a memory debugging tool, such as MemCheck~\cite{Seward:atc2005}, AddressSanitizer~\cite{Serebryany:atc2012}, or DUMA~\cite{duma}, in a production to catch the issue.  But, these tools impose high runtime and memory overhead that are unacceptable for production~\cite{Seward:atc2005}.

\sys{}'s fine-grained interposition enables Charlie to reduce the overhead of their tool.  Charlie determines that the memory corruption is related to objects created by a specific shared library in the program, \texttt{libmpm}.  So, Charlie edits \sys{}'s \relmap{} to use DUMA's interposition only on calls to \texttt{malloc} and \texttt{free} made by \texttt{libmpm}.  Using DUMA on \texttt{libmpm} detects the bug, a buffer overflow of a queue, while minimizing the performance and memory overhead.

\section{Discussion}
\label{sec:discussion}

Today's applications are based upon the premise that symbol resolution is expensive.  Developers place considerable effort into minimizing the number of symbols exported by a dependent library to improve the performance of the applications that use them.  Moreover, lazy binding, a standard optimization that dynamic linkers employ, reduces symbol resolution cost but imposes a per-external function invocation cost.  When using dynamic linking, these efforts and approaches are a useful tradeoff.

However, stable linking offers a pathway to eliminate the presumption of expensive symbol resolution.  Below, we expound upon how it can simplify developer effort surrounding how applications expose symbols (\cref{sec:disc:effort}) and how stable linking can eliminate the need for lazy binding and its per-external function invocation cost (\cref{sec:disc:binding})

\subsection{Reducing Developer Effort}
\label{sec:disc:effort}

Due to the premise that symbol resolution is expensive, we observe few open-source applications with more than \num{200000} symbol relocations. In fact, application and library authors take great pain to minimize their dependency set and the number of symbols that their software exports.  For example, \citeauthor{drepper2006write} describes the effort that shared library writers should take to reduce the number of symbols that they export~\cite{drepper2006write}, and the Android documentation provides similar guidance ~\cite{Controls12:online}.  Today's developers resort to compiler-specific macros to hide symbol visibility, merge their modules into larger compilation units, or change their code entirely to avoid relocations~\cite{Chromium1:online}. 

For example, developers linking the Python bindings for TnFOX into their applications, an extension of the FOX portable GUI toolkit with over \num{250000} symbols, experienced startup times of over six minutes. To address this, the developers augmented GCC to support a visibility attribute that hides many C++ symbols. The approach reduced the number of symbols to under \num{18000} and accelerated startup to only eight seconds.  \sys{} would eliminate the need for this effort since it can quickly load applications with \num{250000} symbols. 

\subsection{Eliminating Lazy Binding}
\label{sec:disc:binding}

The most prevalent approach to minimizing the startup time of symbol relocation is through the use of lazy binding (\cref{sec:motiv:perf}).  Deferring symbol relocations to \emph{first-use} reduces symbol resolution cost, but requires additional data structures that come with additional cost at runtime.  The lazy binding runtime cost has proven significant and spurred the proposal of novel computer architectures~\cite{Agrawal:asplos2015} and post-link compilers~\cite{Ren:osdi2022}.  Since \sys{} accelerates symbol resolution, it offers a simple solution to the overhead of lazy binding: disable it!

\begin{figure}
\begin{minted}[
    linenos,
    frame=single,
    autogobble,
    xleftmargin=1em,
    frame=lines,
    fontsize=\footnotesize,
    linenos=true, 
    escapeinside=!!]{nasm}
main:
  call func@PLT !\tikzmark{A}!

PLT[0]: !\tikzmark{B}!
  call resolver
    
PLT[n]: !\tikzmark{C}!
  jmp *GOT[n] !\tikzmark{D}!
  !\tikzmark{E}!prepare resolver
  jmp PLT[0] !\tikzmark{F}!

GOT[n]:
  !\tikzmark{G}!addr !\tikzmark{H}!

 !\tikzmark{I}!func:
    ...
\end{minted}
\tikzset{Empharrow/.style={dashed,black!60!green,thick,->,arrows={-Triangle[width=3pt]}}}
\tikzset{Memory/.style={purple,thick,->,arrows={-Triangle[width=3pt]}}}
\tikzset{MemoryInit/.style={purple,thick,->,arrows={-Triangle[width=3pt]}}}
\tikzset{MemorySet/.style={blue,thick,->,arrows={-Triangle[width=3pt]}}}
\begin{tikzpicture}[remember picture,overlay]
\draw[Empharrow](A)--+(3cm,0)|-(C)node[pos=0.25,right,text width=4cm]{};
\draw[MemoryInit](F)--+(2.5cm,0)|-(B);
\draw[Empharrow](D)--+(3cm,0)|-(H);
\draw[MemoryInit](G)--+(-0.5cm,0)|-(E);
\draw[MemorySet](G)--+(-0.5cm,0)|-(I);
\end{tikzpicture}
 \caption{At first use, trampoline into dynamic-linker to resolve symbol via the \textcolor{red}{red} arrow. Subsequent resolutions can go straight to the function via the \textcolor{blue}{blue} arrow.}
 \label{fig:plt-symbol-resolution}
 \Description[Symbol resolution via the PLT and GOT.]{Two images that show the trampoline of the PLT into the dynamic linker to resolve symbols.}
\end{figure}
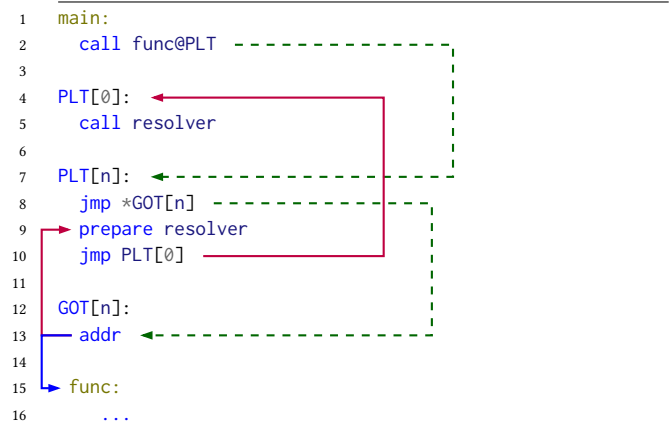

In particular, lazy binding uses two data structures:  the Global Offset Table (GOT), which contains the addresses of global variables and functions, and the Procedure Linkage Table (PLT), which coordinates with the dynamic linker to perform lazy loading.  \Cref{fig:plt-symbol-resolution} shows the PLT and GOT operation.  Each function call to an external symbol in the application calls an entry in the PLT.  Each PLT entry begins with an indirect \texttt{jmp} that uses an entry in the GOT.  Initially, the GOT entry contains an address in the original PLT (the red arrow), which includes logic that prepares and calls the dynamic linker's symbol resolver.  The dynamic linker updates the GOT entry to point to the correct location in the program, so that future invocations skip the resolver path (the blue arrow).  This double \texttt{jmp} is often referred to as a \emph{trampoline} and the source of the penalty incurred by lazy binding throughout the application's execution each time a function is called.

\begin{figure}
\centering
\includegraphics[width=.9\columnwidth]{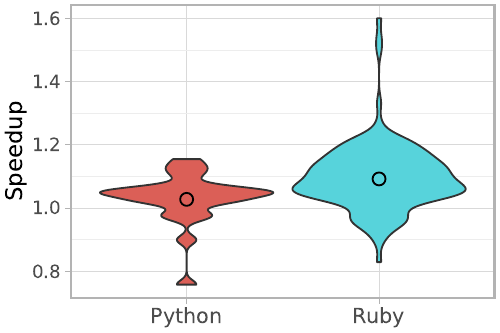}
\caption{Violin plots of the speedup of Python and Ruby across their benchmark suite when omitting the PLT. The circle shows the geometric mean speedup.}
\label{fig:benchmark-python-ruby-plt}
\Description[Speedup of Python and Ruby bench suite when built without PLT.]{Speedup of Python and Ruby bench suite when built without PLT.}
\end{figure}

Our solution disables the PLT, while leaving the GOT in place.  Thus, it eliminates a \texttt{jmp} instruction for each invocation of an external function.  We built the Ruby and Python interpreters with and without a PLT and execute the complete Ruby and Python benchmark suites~\cite{RubyBenc94:online, pythonpy81:online} on each version.   \Cref{fig:benchmark-python-ruby-plt} shows violin plots of the speedup from removing the PLT compared to using it for both benchmarks.  The geometric mean of the speedup for ruby and python is $9.22\%$ and $2.75\%$, respectively.  This illustrates the potential knock-on effects of a faster symbol relocation process.

\section{Related Work}
\label{sec:rel}
\label{sec:related}











Stable linking improves the performance and dependency management of dynamic linking, while \sys{} is the first stable linker.  We describe prior work on optimizing dynamic linking, optimizing application startup, and works that enable dependency management.

\paragraph{Dynamic Linking Optimizations}
The cost of dynamic linking is well studied~\cite{drepper2006write} and has led to the development of many optimizations.  Unfortunately, existing solutions either forgo the benefits of dynamic linking, do not effectively reduce dynamic linking costs, or are incompatible with standard system designs.  In particular many works optimize dynamic linking internals~\cite{PATCHDTG46:online, specification1993tool, drepper2006write, direct-binding:online,direct-binding:online, dt-gnu-hash:online}
and retain many of its performance limitations.  Shrinkwrap~\cite{Zakaria:hpc2022} accelerates dynamic linking's startup cost by caching the libraries that will make up an applications' dependencies, but is less effective than \sys{} since it still performs symbol relocations online.  Some approaches eliminate the dynamic linking runtime overhead by overhauling the linking approach but are incompatible with standard features of current systems. For example, Prelink~\cite{Jung:emsoft2007, prelink:online} is incompatible with ASLR, while Software Multiplexing~\cite{Dietz:oopsla2018} is incompatible with current compilers and binary-based software distribution.  Finally, some systems use static linking (e.g., Slinky~\cite{Collberg:atc2005}) and thus improve startup performance but lose the code reuse, reduced network/disk use, and streamlined software updates of shared libraries.

Some systems focus on other challenges with dynamic linking rather than startup overhead; \sys{} is orthogonal and comparable with these efforts.  For example, \citeauthor{Agrawal:asplos2015} propose a speculative hardware mechanism, to accelerate dynamic linking trampolines created due to lazy binding~\cite{Agrawal:asplos2015}, while iFed provides a post-link optimization pass for eliding some compiler-created trampolines~\cite{Ren:osdi2022}.  \sys{} is compatible with such approaches; although, we show that it is feasible to turn of lazy binding altogether and observe similar performance benefits (\cref{sec:disc:binding}).  Guided linking provides a mechanism for performing compiler optimizations over an application and its dynamically loaded libraries by narrowing the possible dynamic libraries that can be loaded at runtime~\cite{Bartell:oopsla2020}.  Stable linking empowers the same optimization.

\paragraph{Application Startup Optimizations}

Many systems accelerate application start-up, but none target symbol relocation.  For example, many systems accelerate library loading by optimizing storage speeds for linux (e.g., FAST~\cite{Joo:fast2011}), windows (e.g., TurboMemory~\cite{Mathews:tos2008}) and on mobile devices (e.g., FALCON~\cite{Yan:mobisys2012}, CAS~\cite{Lee:ubicom2016}, and APPM~\cite{Parate:ubicom2013}).  Other efforts target application startup for specific applications such as Java~\cite{Lion:osdi2016}.  Shortcut~\cite{Dou:sosp2019} accelerates mostly-deterministic regions of an execution, such as application startup, by memoizing parts of a program that are likely to remain constant across executions.  Shortcut's startup improvements would likely benefit from stable linking as it creates additional determinism across application startup. 

\paragraph{Dependency Management}


There are few works that aid with observing and updating dependencies as needed for auditing, debugging, and updating software systems.  For example, \texttt{llvm-ifs}~\cite{llvmifss19:online} and \texttt{readelf}~\cite{readelf111:online} extract ABI-related information in a human-readable format, but do not identify how the system will resolve symbols at runtime.  Consequently, most dependency management today relies on system administrator best practices\cite{Howtoche20:online}.


\section{Conclusion}
\label{sec:conclusion}

We presented stable linking, a new approach to using external dependencies that provides the benefits of shared libraries---code reuse, reduced network/disk use, and streamlined software updates---while resolving the principle limitations of dynamic linking---poor startup performance and poor dependency management.  A stable linked system can be in one of two states: management time, in which its dependencies and applications are modifiable, and epochs, in which its dependencies and applications cannot change.  We then described \sys{}, the first stable linker.  \sys{} materializes each application's relocations at the end of each management time.  It loads these relocations during execution; we show that this approach accelerates real world benchmarks by a factor of  \latencySpeedupGeo{} on average.  \sys{} exposes the relocations to users to improve dependency management; we show the utility of this approach by using it in \numvignettes{} vignettes that evaluate ABI compatibility, audit a system's susceptibility to a new CVE, and debug a memory issue. In presenting \sys{}, we bridge the gap between the benefits of shared libraries and the performance challenges of dynamic linking.


\bibliographystyle{plainnat}
\bibliography{references}

\end{document}